\documentclass[acmsmall,screen]{acmart}

\AtBeginDocument{%
  \providecommand\BibTeX{{%
    \normalfont B\kern-0.5em{\scshape i\kern-0.25em b}\kern-0.8em\TeX}}}

\setcopyright{acmcopyright}
\copyrightyear{2022}
\acmYear{2022}
\acmDOI{XXXXXXX.XXXXXXX}

\acmJournal{tosem}

\usepackage{xspace}
\usepackage{amsmath,amsfonts}
\usepackage{algorithmic}
\usepackage{cleveref}
\usepackage{graphicx}
\usepackage{textcomp}
\usepackage{enumitem}
\usepackage{multirow}
\usepackage{booktabs}
\usepackage{subcaption}
\usepackage{tcolorbox}
\usepackage{threeparttable}
\usepackage{xcolor}
\usepackage{soul,balance}
\usepackage{listings}
\usepackage{colortbl}
\usepackage{hhline}

\newcommand{\su}[1]{{\color{black}#1}}

\newcommand{\mechanism}{FrugalCoder\xspace}
\newcommand{\estimator}{TCQE\xspace}

\begin{document}

\title[Preventing Unhelpful Code Completion for Productive and Sustainable Neural Code Completion Systems]{Don't Complete It! Preventing Unhelpful Code Completion for Productive and Sustainable Neural Code Completion Systems}

\author{Zhensu Sun}
\affiliation{%
  \institution{Singapore Management University}
  \country{Singapore}
}
\email{zssun@smu.edu.sg}

\author{Xiaoning Du}
\affiliation{%
  \institution{Monash University}
  \city{Melbourne}
  \state{Victoria}
  \country{Australia}
}
\email{xiaoning.du@monash.edu}
\authornote{Corresponding author.}

\author{Fu Song}
\affiliation{%
  \institution{Key Laboratory of System Software (Chinese Academy of Sciences), State Key Laboratory of Computer Science, Institute of Software, Chinese Academy of Sciences}
  \city{Beijing}
  \country{China}}
\email{songfu@ios.ac.cn}
  \additionalaffiliation{
\institution{University of Chinese Academy of Sciences}
  \city{Beijing}
  \country{China}}
\additionalaffiliation{
\institution{Nanjing Institute of Software Technology}
  \city{Nanjing}
  \country{China}}

\author{Shangwen Wang}
\affiliation{
  \institution{National University of Defense
Technology}
  \city{Changsha}
  \country{China}
}
\email{wangshangwen13@nudt.edu.cn}

\author{Mingze Ni}
\affiliation{%
  \institution{University of Technology Sydney}
  \city{Sydney}
  \state{New South Wales}
  \country{Australia}
}
\email{Mingze.Ni@student.uts.edu.au}

\author{Li Li}
\affiliation{%
  \institution{Beihang University, Beijing}
  \city{Yunnan Key Laboratory of Software Engineering}
  \country{China}
}
\email{lilicoding@ieee.org}

\author{David Lo}
\affiliation{%
  \institution{Singapore Management University}
  \country{Singapore}
}
\email{davidlo@smu.edu.sg}

\begin{abstract}
Currently, large pre-trained language models are widely applied in neural code completion systems.
Though large code models significantly outperform their smaller counterparts, around 70\% of displayed code completions from Github Copilot are not accepted by developers.
Being reviewed but not accepted, their help to developer productivity is considerably limited and 
may conversely aggravate the workload of developers, as
the code completions are automatically and actively generated in state-of-the-art code completion systems as developers type out once the service is enabled.
Even worse, considering the high cost of the large code models, it is a huge waste of computing resources and energy, which severely goes against the sustainable development principle of AI technologies.
However, such waste has never been realized, not to mention effectively addressed, in the research community for neural code completion.
Hence, preventing such unhelpful code completions from happening in a cost-friendly way is of urgent need.
To fill this significant gap, we first investigate the prompts of unhelpful code completions, called ``low-return prompts''.
We empirically identify four observable patterns in low-return prompts, each lacking necessary information, making it difficult to address through enhancements to the model's accuracy alone.
This demonstrates the feasibility of identifying such low-return prompts based on the prompts themselves.
Motivated by this finding, we propose an early-rejection mechanism to turn down low-return prompts by foretelling the code completion qualities.
The prompts that are estimated to receive unhelpful code completions will not be sent to the model.
Furthermore, we investigated five types of estimators to demonstrate the feasibility of the mechanism.
The experimental results show that the estimator can reject 20\% of code completion requests with a 97.4\% Precision.
To the best of our knowledge, it is the first systemic approach to address the problem of unhelpful code completions and this work also sheds light on an important research direction of large code models.
\end{abstract}



\maketitle

\section{Introduction}

Benefiting from the pre-trained Large Code Models (LCMs), automated code completion systems powered by deep learning have achieved unprecedented superior performance, and thus have great potential to significantly lower the barrier of entry for programming and increase developer productivity.
Motivated by the strong product-market fit and immense business value, a large number of commercial applications based on LCMs are recently released,
including Github Copilot~\cite{copilot}, aiXcoder~\cite{aixcoder},
TabNine~\cite{tabnine},
and CodeWhisperer~\cite{codewhisperer}.
LCMs in practice contain billions of parameters, and consequently, have to be deployed on high-performance servers and supply service via remote APIs.
Aiming for seamless assistance to developers, code completion systems are designed to automatically and actively issue code completion requests to the servers when a typing phase is detected.
To the best of our knowledge, users are not enabled to configure how frequently and when the requests are issued to the code completion services.
Every request activates the LCM to generate a piece of code completion that will be displayed to the user.
According to a survey~\cite{Ziegler2022ProductivityAO} on GitHub Copilot, an average number of 22.5 code completions are displayed to a user per hour.

However, not all the code completions generated by the state-of-the-art code completion systems are helpful to the developers.
For instance, according to the study of Ziegler et al.~\cite{Ziegler2022ProductivityAO} where they surveyed developers' feedback on Copilot code completions and 2,631 responses were collected,
around 70\% of displayed code completions are \emph{not} accepted by developers.
Undoubtedly, such a large proportion of unhelpful code completions pose significant threats to the main objective of code completion systems -- to improve developer productivity, and put relevant operation costs in vain.
On the one hand, even if a code completion is not accepted by the developer, it often requires substantial effort to be carefully (sometimes roughly) reviewed, incurring negative impact on the developers' productivity.
Some recent studies~\cite{vaithilingam2022expectation,Barke2022GroundedCH} found
that developers might get rid of the entire code completions and start over the coding themselves.
On the other hand, generating these unhelpful completions is a waste of computing resources and energy,
which severely goes against the sustainable development principle of AI technologies~\cite{Wynsberghe2021SustainableAA}.
In particular, in addition to training, larger-scale models burn more resources and energy during daily inference tasks after deployment~\cite{Desislavov2021ComputeAE}, thus the (wasted) computational cost from unhelpful code completions can be massive.
Furthermore, in practice, to fulfill the enormous user requests in a timely manner, code completion servers usually employ advanced concurrency techniques to achieve high throughput, which also increases their operation costs.
In summary, both the harms to developer productivity and computational wastes brought by unhelpful completions in neural code completion systems are unacceptable, and should be avoided, when possible,
via building more productive and sustainable code completion systems.

To reduce unhelpful code completions, one first has to understand the cause of unhelpful code completions.
In general, we found there are two main possible reasons: the model is not doing a good job (due to a lack of, e.g., high-quality training datasets, good-enough neural networks, or training approaches)
or \su{the prompts are not informative enough to derive proper completions}.
The former may be fixed by optimizing LCMs themselves as technology advances, while the latter is \su{hard to solve if no additional measures are taken.}
For example, it is almost impossible to initialize a variable named with meaningless characters (e.g., \emph{abc}) unless more information is given.
In fact, we can hardly expect an omnipotent LCM that avoids all unhelpful completions since there are always out-of-capability and \su{uninformative} prompts.
Hence, to explore whether and how these unhelpful code completions can be prevented from happening is of urgent need, and to the best of our knowledge, this problem has never been realized, not to mention effectively addressed, in the research community for neural code completion.

To fill the above gap, we first investigate the feasibility of recognizing \emph{low-return} prompts, i.e., the prompts leading to unhelpful code completions, and then propose a novel and cost-friendly solution to preclude general low-return prompts before forwarding them to the completion server.
Our approach not only benefits developers' productivity by freeing them from reviewing some unhelpful code completions,
but also saves the computational cost of LCM servers for sustainable development.
The feasibility mainly relies on the existence of recognizable patterns among low-return prompts, such that rule-based or learning-based algorithms could be devised to identify them.
Therefore, a manual inspection to verify the existence of such patterns is necessary.
While the prompts of these unaccepted code completions reported in~\cite{Ziegler2022ProductivityAO} form an ideal subject for the inspection,
they are currently not publicly available.
Alternatively, we crafted a small dataset where each entry contains a prompt, its ground truth completion, and the completion given by Github Copilot.
Two authors manually labeled if the completion given by Github Copilot is helpful with a reference to its ground truth, and examined whether some patterns that lead to low-return prompts exist in prompts whose inferred completions are unhelpful.
Finally, we empirically identified four low-return prompt patterns.
This underpins the feasibility of our idea for precluding the low-return prompts.

\begin{figure}[t]
\centerline{\includegraphics[width=0.98\linewidth]{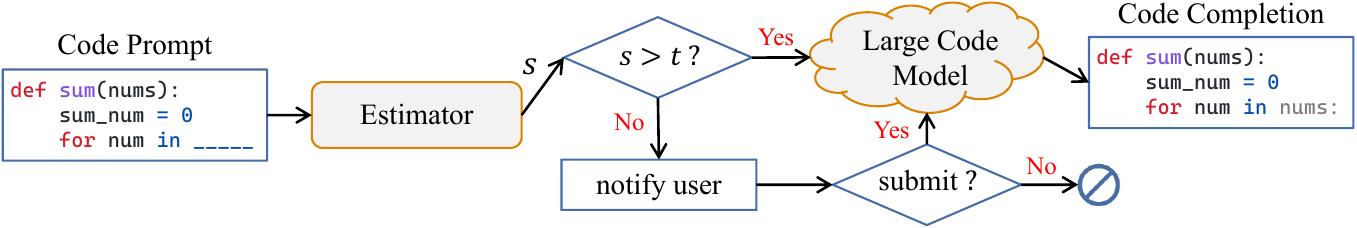}}
\caption{The workflow of an LCM-based code completion system with \mechanism\su{: Given a code prompt, the estimator first produces an estimation score $s$ that reflects the anticipated quality of completion. If $s$ exceeds a predefined threshold $t$, the LCM is activated to fulfill the completion request. Conversely, if $s$ falls below $t$, the request for completion is denied.}}
\label{fig:overview}
\end{figure}

Based on the above findings, we propose to equip the LCM-based code completion systems with an early-rejection mechanism to turn down low-return prompts, named \mechanism.
We estimate the quality of completions since it is a quantifiable indicator of the abstract concept—— helpfulness, where low-quality completions are more likely to be unhelpful.
\mechanism is designed to build a productive and sustainable system for LCM service providers, e.g. Github Copilot, for whom the operation cost and developer productivity~\cite{kochhar2015understanding} are critical to their business.
As shown in~\Cref{fig:overview}, the core idea of \mechanism is to have a lightweight code completion quality estimator for the code completion system, which guards the prompts sent to the LCM.
It foretells the completion qualities of the LCM solely based on prompts. 
Whenever the estimated completion quality is lower than a chosen threshold, it blocks the completion request.
To guarantee developer productivity and computing cost reduction simultaneously, the mechanism must be effective and cost-friendly.
The former requires the estimator to accurately estimate the completion quality for prompts, especially the low-return prompts, and the latter is to create and utilize the estimator in a cost-friendly way, such that the overall cost of the code completion system can be significantly reduced.
As a newly identified research problem in the context of neural code completion systems, how to design such an estimator is yet to be investigated.

There exists a trade-off between the effectiveness and cost-friendliness of a code completion quality estimator.
The optimum completion quality estimator is the LCM itself, however, no energy could be saved in this way.
With a lighter but less accurate estimator, it may underestimate or overestimate the actual completion qualities, where the former may hinder the usability of the
code completion system and the latter may leave some low-return prompts unattended.
Thus, a systematic investigation of the estimator design is valuable but challenging.
Manually designing an accurate rule-based completion quality estimator is possible in theory, but it is labor-intensive and hard to be complete in practice.
\su{
In this work, we look into learning-based estimators, and, for the first time, implements and assess such estimators.
To better understand the technical design of these estimators, we implement multiple estimators with different learning-based methods, including, Linear Regression, Adaboost, LSTM, and Transformers (both encoder-only and decoder-only).
With these estimators, we aim to demonstrate the feasibility of \mechanism, and assess how well such an estimator can fulfill the goals of \mechanism.
}

\su{
To sufficiently evaluate the effectiveness of these estimators and the feasibility of \mechanism, we conduct a comprehensive evaluation, involving two programming languages (Java and Python), two LCMs (CodeGen2 and StarCoder), and two metrics (BLEU and CrystalBLEU).
Moreover, we build a human-annotated benchmark, HumanAccept, where the model-generated code completions are annotated with whether they will be accepted or not.
With the benchmark, we first measure whether each estimator can accurately block low-return prompts under different threshold settings.
Among the implemented estimators, encoder-only Transformer is the best-performing one and can achieve a 97.4\% Precision to reject 20\% of code prompts.
Furthermore, we investigate the developer productivity of LCMs with/without \mechanism, where the average acceptance rate of displayed code completions can rise from 27.4\% to 33.0\% if 20\% of code prompts are rejected by the encoder-only Transformer.
More importantly, given a code prompt, it only takes less than 5.2ms for these estimators to yield an estimation score, while CodeGen2 and StarCoder take 3.9s and 1.7s respectively to generate the completion (containing 10 tokens).
}

To the best of our knowledge, we are the first to propose a general and effective mechanism to prevent unhelpful code completions in a cost-friendly manner for LCM-based code completion systems.
Our contributions are summarized as follows:
\begin{itemize}
    \item A preliminary study that reveals the existence of the observable patterns in low-return prompts, which demonstrates the feasibility of inventing an effective recognizer.
    \item A simple yet effective prompt early-rejection mechanism, \mechanism, to improve the developer productivity and save the computational costs of LCM-based code completion systems.
    \item \su{A range of estimators implemented for \mechanism, which can accurately estimate the performance of LCMs on a given code snippet in a cost-friendly manner.}
    \item A comprehensive evaluation that demonstrates the feasibility of \mechanism by evaluating the effectiveness and cost-friendliness of the estimators.
\end{itemize}

The source code, artifacts and results of \estimator are released to facilitate future research and industrial practices, which are available at~\url{https://github.com/v587su/FrugalCoder}.

\smallskip
\noindent
{\bf Organization.}
The rest of the paper is organized as follows. \Cref{sec:2} prepares readers with the definitions and patterns of unhelpful code completions and low-return prompts. \Cref{sec:3} describes the design of our mechanism to turn down low-return prompts. \Cref{sec:DCE} introduces the proposed estimators for the mechanism. \Cref{sec:setup} shows the experimental settings and the evaluation results are discussed in~\Cref{sec:results}. \Cref{sec:threats} discusses the potential threats to the validity, potential applications, and limitations. Finally, \Cref{sec:related} introduces related work and \Cref{sec:conclusion} concludes the paper.

\section{Unhelpful code completions and Low-return prompts}
\label{sec:2}
In this section, we clarify the important terminologies used throughout the article, namely, unhelpful code completions and low-return prompts.
Basically, low-return prompts can lead to unhelpful code completions.
Our work exactly aims at turning down low-return prompts to prevent unhelpful completions without feeding the prompts to the LCM.
As enlightenment to our idea, in a manual investigation of these low-return prompts, we empirically summarized four recognizable patterns of low-return prompts.

\subsection{Unhelpful Code Completions}

Being unhelpful means that the completions generated by code completion systems fail to boost, sometimes even inhibit, developers' productivity, for example, when they do not match developers' intention or demand non-negligible efforts for understanding and/or debugging.
Intuitively, in a scenario where a piece of code completion is generated by the system according to an input prompt, there are two main possible reasons for the ineffectiveness of the completion, notably,
the system is not doing a good job or the prompt itself is not informative enough to derive a proper recommendation.
Prompts beyond the capability or \su{being uninformative} (such as prompts without clear intentions) are difficult, or even impossible, to fix by optimizing LCMs.
In fact, we can hardly have an omnipotent LCM, hence unhelpful code completions always exist and must be countered due to their harmfulness to productivity and unacceptable computational waste.

Researchers have been looking into the helpfulness of Github Copilot ever since its birth, by surveying user feedback on their experience with Copilot~\cite{vaithilingam2022expectation,Barke2022GroundedCH} or collecting statistics about the portion of Copilot completions that get discarded without adoption~\cite{Ziegler2022ProductivityAO}.
The human study in~\cite{vaithilingam2022expectation} revealed that three out of the five cases, where participants failed to accomplish the programming task, were caused by the misleading recommendations from Copilot.
According to another study~\cite{Ziegler2022ProductivityAO}, among the Copilot completions displayed to users, around 70\% were not accepted.
Such unhelpful completions waste both the developers' effort to review them and the valuable computational resources and energy of the code completion systems.
This work intends to prevent such completions from happening and raises attention to the cost-effectiveness of designed solutions.

\subsection{Low-return Prompts}
\label{sec:prompts}

The prompts that lead to unhelpful code completions are called low-return prompts in this paper.
If there is a way to recognize these low-return prompts in advance and preclude them before igniting LCMs for inference, the drawbacks coming along with unhelpful completions can be avoided.
The recognizer must effectively distinguish such low-return prompts and work in a cost-friendly manner.
Its feasibility highly depends on whether such low-return prompts share some recognizable characteristics in common, such that they could be captured with either rule-based or learning-based algorithms.

To this end, we conduct a small-scale empirical study and attempt to extract patterns of low-return prompts, if any, through manual inspection.
The desired subjects are the production data of the unaccepted completions as studied in~\cite{Ziegler2022ProductivityAO}, but they are not publicly accessible.
As an alternative, we crafted a group of prompts in pair with their corresponding ground truth code completions by cutting code snippets into two parts at random positions, where the prefix works as the prompt and the suffix is the expected completion.
The code snippets are randomly sampled from a Java code dataset, COFIC~\cite{Sun2022OnTI} and more details of this dataset are presented in \Cref{sec:setup}.
\su{The sample size~\cite{Krejcie1970DeterminingSS} is statistically decided by:
\begin{equation}\label{eq:1}
  ss = \frac{z^2 \times p\times(1-p)/c^2}{1+\frac{z^2\times p\times(1-p)/(c^2-1)}{population}}
\end{equation}
where $population$ is the size of the group (set to the size of the entire test dataset for adequate sampling), $p$ is the standard deviation of the population, $c$ is the confidence interval (margin of error), $z$ is the Z-Score determined by the confidence level.
In this experiment, we choose to work with a 95\% confidence level (i.e., 1.96 Z-Score according to \cite{Krejcie1970DeterminingSS}), a standard deviation of 0.5 (0.5 is the maximum standard deviation, which is a safe choice given the exact figure unknown), and a confidence interval of 5\%.}
Finally, we obtained 383 prompt-and-ground-truth pairs.
The prompts are fed into a commercial code completion application, Github Copilot, to get the completion recommendations.
Note that Github Copilot automatically triggers a completion request when the coding suspends for a while, so that every prompt obtains one completion (including the blank completion).

\begin{figure}[t]
\centerline{\includegraphics[width=0.55\columnwidth]{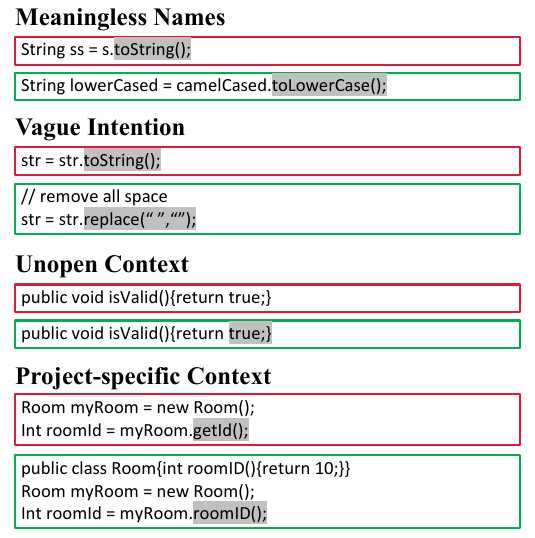}}
\caption{The simple examples of the prompts following each summarized pattern. The example following the pattern is in red, and the slightly shaped variant is in green. The completions are marked with the gray background.}
\label{fig:examples}
\end{figure}

Two authors were appointed to separately decide whether a code completion is helpful.
\su{
More specifically, given the code context, generated code completion, and the ground truth, they are asked to mark if the generated code completion can be accepted as a continuation of the code context.
The continuation should exactly match or achieve the same intention as the ground truth (the next ten tokens in the origin code snippet).
The ground truth is provided to help annotators accurately understand the exact intent of the code context.
Therefore, syntactically correct but unintended completions are also labeled as unaccepted. 
The Cohen’s Kappa agreement levels between the two authors are 0.77, revealing substantial agreement between the two authors.
Afterward, they had a discussion to reconcile all disagreements.
Finally, 200 out of 383 completions are considered as unhelpful.

According to the observed unhelpful completions, they further examined the corresponding prompts, i.e., the low-return prompts.
The examination is to empirically find out the root cause for the prompt being low-return.
To be specific, the two authors reviewed each unhelpful completion together and discussed what patterns in its prompt could lead to the unhelpful completions based on their domain knowledge, where the identified patterns are incrementally added to a list. 
Please note that we don't seek to find a complete list of the features of low-return prompts, but just a demonstration to exemplify the existence of such characteristics.
Therefore, from the list of patterns, we keep the four most apparent patterns: Meaningless Names, Vague Intention, Unopen Context, and Project-specific Context, which appeared in 94, 155, 4, and 180 low-return prompts, respectively.}
We now present the details of the four patterns as follows.
For ease of understanding, as listed in~\Cref{fig:examples}, each pattern is accompanied by a simple prompt example following that pattern, and a slightly shaped variant of that prompt to be an opposite pattern.

\begin{itemize}[leftmargin=*]
\item \textbf{Meaningless Names}:
Identifier names, such as variable names and method names, play an important role in completing code prompts, either for human developers or automated tools.
Java naming convention~\cite{javaConvention} also recommends using mnemonic names which indicate the intent of their usage to casual observers.
Meaningless names offer little information about the type, intention, or other critical properties of the identifiers to the code completion systems, making it difficult, sometimes even impossible, to strike completions desired by the developers.
As shown in the example, when to get the lower-cased version of a string, the variable name \emph{lowerCased} provides a much clearer hint, than \emph{ss}, to the inference model on recommending the \emph{toLowerCase()} API.

\item \textbf{Vague Intention}:
In addition to identifier names, code intention can also be conveyed through comments.
When the intention is not indicated in either way, completion models can hardly generate a beneficial completion for a prompt, since there are too many plausible options.
The example demonstrates a case where the developer intends to remove spaces in a string.
When neither meaningful names nor comments are presented to reveal the coding intention, the completion model tends to find a popular String API instead of going directly to the expected one.

\item \textbf{Unopen Context}:
For prompts ending with enclosed code structures, e.g., finished class or function definitions, little can be done on the completer side, since the code development has come to an intermission.
Even if the completion model properly generates a blank completion for such prompts, this is still unnecessary consumption of energy and computing resources.
The example shows a prompt of an enclosed function definition, where no completions are needed.

\item \textbf{Project-specific Context}:
As opposed to built-in libraries and popular third-party libraries of a programming language, project-specific code, such as global variables, self-defined functions, and customized classes, is rarely seen when training code completion models.
If completing a prompt relies on the understanding of some project-specific code, the reasoning is difficult to succeed if such information is missing from the prompt.
In the example, the class \emph{myRoom} is a self-defined Java class, and it would be hard to recommend the invocation to \emph{roomID()} without knowing the list of methods defined in it.

\end{itemize}

\section{Frugal Code Completion}
\label{sec:3}
Currently, most of the low-return prompts are indiscriminately handled by LCM-based code completion systems, which leads to enormous unreliable suggestions and resource waste.
To improve productivity and save costs, we look into this new research topic: \textbf{how to selectively handle code completion requests in a cost-friendly manner without significantly affecting the user experience with the system}.

\subsection{Problem Setup}
Our empirical study in \Cref{sec:prompts} illustrates the feasibility of designing an effective recognizer for such low-return prompts by identifying their observable patterns.
With such a recognizer, we can block the recognized low-return prompts out from the code completion systems, which avoids the generation of unhelpful completions.
Therefore, without loss of generality, we propose to have a lightweight completion quality estimator which foretells how well a prompt can be completed by an LCM without executing the actual inference and call the mechanism \mechanism.
Various methods can be adopted to implement the estimator\su{, including heuristic rules and learning-based models.}
Prompts will be turned down without getting completed if the completion quality is estimated to be lower than a pre-defined threshold.
The strictness of \mechanism can be flexibly controlled by adjusting the threshold.
To be specific, we define \mechanism as follows:
\begin{tcolorbox}[size=title]
\textbf{Definition 1}

Given an LCM-based code completion system $M$ and a code completion prompt $p$, a completion quality estimator $E$ will estimate the quality of the completions to be generated by $M$ for $p$, i.e., before $p$ is actually processed.
If the estimated score $s$ is lower than a pre-defined threshold $t$, $p$ will be blocked from being sent to the LCM-based code completion system $M$.
\end{tcolorbox}

\su{The LCM-based code completion system here refers to the system that predicts the subsequent tokens of a given code prompt and shows the suggestions to its user.
We demonstrate the workflow of an LCM-based code completion system equipped with \mechanism in \Cref{fig:overview}.
Under \mechanism, each code completion prompts sent to the system will be first measured by the estimator.}
Whenever the estimated completion quality is less than a pre-set threshold, the prompt is blocked and users will be notified.
They can choose to follow the notification -- give up sending a completion request for the current prompt and continue optimizing it to be informative, or ignore the notification and issue the completion request anyway.

\subsection{Technical Challenges}
\label{sec:challenges}
The effectiveness of \mechanism depends on how many unhelpful completions can be prevented and how much wasted energy can be saved.
Ideally, \mechanism is expected to block all of the low-return prompts perfectly,
however, it is almost impossible to implement such a \mechanism.
In practice, according to the preference of the LCM service providers, \mechanism can be applied as long as it can effectively improve productivity and/or save computing costs.
More specifically, \mechanism seeks to achieve higher accuracy in identifying low-return prompts and lower computing consumption, which mainly depend upon the performance of the estimator, and the challenge is to optimize both the accuracy and the cost-friendliness.

\smallskip
\noindent\textbf{Accuracy of the estimator.}
The estimator should be accurate, i.e., estimating as close as possible to the actual completion qualities,
as estimators with poor accuracy will 
harm the productivity brought by auto-completions and limit the gain of resource saving.
For example, wrongly predicting a prompt as low-return hinders developers' access to a potentially helpful completion, and they may spend extra efforts to re-submit the completion request. 
However, as an emerging problem, the estimator is not yet well investigated by the research community.
It is still unclear how to design and implement an effective estimator that can accurately estimate the completion quality of the LCM given a code prompt, which calls for more attention from the research community.

\smallskip
\noindent\textbf{Cost-friendliness of the estimator.}
Considering one of the goals of \mechanism, i.e., saving the cost of LCM-based code completion systems, the estimator is required to be cost-friendly in computing.
The cost of the estimator itself should be considered when calculating the energy and resources saved with \mechanism.
Thus, we cannot infinitely enlarge the scale of the estimator to obtain more accurate estimations on completion qualities.
How to balance the trade-off between the effectiveness and efficiency of the estimator requires further exploration.

\section{Learning-based Completion Quality Estimator}
\label{sec:DCE}

\begin{figure*}[t]
\centerline{\includegraphics[width=1\textwidth]{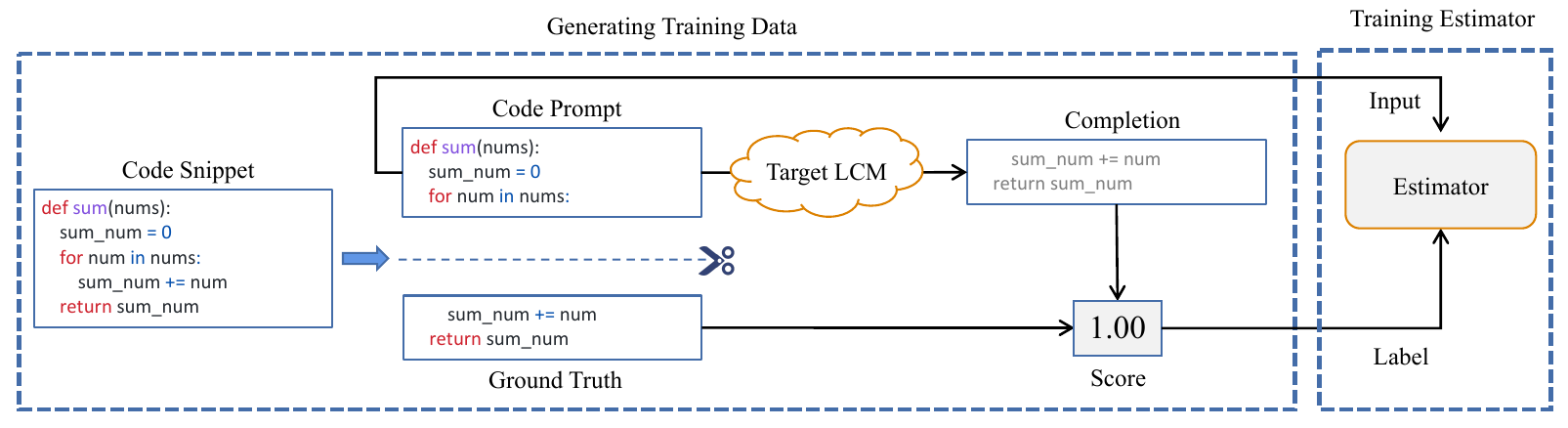}}
\caption{The training process of a learning-based estimator.}
\label{fig:train}
\end{figure*}

The estimator is the core of \mechanism.
\su{As presented in~\Cref{sec:prompts}, the low-return prompts often follow complicated patterns, which are hard to be covered by the rule-based heuristics.
Consequently, an estimator capable of automatically understanding the code semantics of these prompts is preferable.
Considering the outstanding performance of learning-based methods in semantic understanding of source code, we seek to train a model to map between the prompts (in the training dataset) and the quality of their completions from a target LCM, which can be seen as a regression task.
More specifically, we propose a methodology to train various learning-based models, transforming them into completion quality estimators.
}

Intuitively, code completions are less likely to be helpful to developers if they are of low quality.
However, out of privacy and business considerations, there is no public dataset on the quality profile computed from user behaviors, such as acceptance rate, by in-production LCM-based code completion systems.
Thus, in this work, we resort to accuracy-based metrics, e.g., BLEU, as replacements for measuring the quality of code completions.
\su{These metrics are designed to evaluate the precision of model-generated completions against a known correct completion, i.e., the ground truth.
Conveniently, these accuracy scores can be directly obtained from the target LCM with general code datasets, significantly expanding the training dataset for our estimator.
In the subsequent sections, we detail the methodology for preparing the training dataset for the estimator and discuss the selection of technical models.}


\subsection{Training Dataset Preparation}
\label{sec:generating_dataset}
\su{To craft the training dataset for a learning-based estimator, each data point should include a code prompt alongside the corresponding accuracy score of its completion (generated by a target LCM).
The accuracy score is computed between the model-generated completion and its ground truth completion.
To obtain the necessary code prompt and its ground truth completions, we split a piece of code $C$ at a random position into two segments.
The former segment $C_{x}$ works as the prompt, while the latter segment  $C_{y}$ serves as its ground truth completion.
Notably, we can reuse the training dataset of the target LCM for the code dataset, which is typically collected from open-source repositories.
Subsequently, the target LCM, $M$, is employed to generate a completion prediction, $C_{y}'$, for the prompt $C_{x}$.
The correctness of this prediction to the actual completion, $C_{y}$, is assessed using a predefined accuracy metric, resulting in a score, $s$.}
We remark that for the prompt $C_x$ whose corresponding ground truth $C_{y}$ is null, i.e., the prompt is in a finished status with unopen context, we directly set its corresponding score to the minimum value instead of querying the target LCM.
It enables the estimator to recognize prompts in the \emph{Unopen Context} pattern as low-return.
Finally, by pairing each prompt $C_x$ with its score $s$, we obtain a code-score dataset, with which we train the estimator model.


\su{Notably, in this method, we focus on single-line level code completion, instead of multiple-line code completions.
Using multiple-line code completions might inadvertently degrade the quality assessment of single-line completions due to the drawbacks of exact-match-based metrics.
For example, given a piece of code completion with an erroneous initial line and correct rest lines, a multiple-line estimator will pass its request for the high exact-match score.
In contrast, a single-line estimator can properly handle this case if applied line by line. 
Thus, we kept the first $k$ tokens ($k=10$ in our experiments) of each ground truth completion $C_{y}$ for a full-line code completion suggestion.
}

\subsection{Estimator Models}

\label{sec:estimator_design}
\su{Various models are capable of performing regression tasks, each with its unique advantages.
Considering the diverse deployment environments and requirements in practice, we argue that no single model can serve as a universal solution for the estimators of \mechanism.
Therefore, we implemented multiple estimators with different learning-based techniques, encompassing both machine learning and deep learning, to assess \mechanism's feasibility and to highlight specific application scenarios of each technique.
In total, we design and implement five estimators using three common learning-based approaches.

\smallskip
\noindent \textbf{Transformer}: Transformer~\cite{Vaswani2017AttentionIA} has shown outstanding performance on various code understanding tasks~\cite{Chen2021EvaluatingLL}.
We implemented two Transformer-based models: one following BERT's Encoder-only architecture~\cite{Devlin2019BERTPO} (\textbf{ET}) and another based on GPT-2's Decoder-only architecture~\cite{Radford2019LanguageMA} (\textbf{DT}).
These models include a linear head layer for regression, mapping the backbone's hidden state to an estimated accuracy score.
Aiming for a lightweight Transformer design, we minimized the hyper-parameters of the models, where each Transformer model is set to have one hidden layer and four attention heads.
We leave the embedding size of them unchanged so that ET and DT can respectively reuse the embedding layers from BERT and GPT-2, which enhances the estimator's semantic understanding.
Notably, though the embedding layer from pre-trained models enlarges the size of the estimator, it won't significantly increase the inference cost as the embedding layer merely stores the embedding vector of each token, where no matrix multiplication is required.
Given an input code prompt, the estimator begins by tokenizing the prompt using the Byte-Pair Encoding (BPE) tokenizer~\cite{Sennrich2016NeuralMT}, retrieving their embedding vectors from the embedding layer and then proceeds to predict the score. 
Both ET and DT are trained to minimize the Mean Squared Error (MSE) between the predicted and actual scores, calculated over the training samples.

\smallskip \noindent \textbf{Recurrent Neural Network (RNN)}: 
The RNN, particularly its Long Short-Term Memory (LSTM) variant, is widely applied in many deep learning tasks, such as code search~\cite{gu2018deep} and code comment generation~\cite{Hu2018DeepCC}, before the emergence of the Transformer.
Recognizing its efficiency in handling sequential data such as code, we have implemented an RNN-based estimator, which is designed similarly to our Transformer-based approach.
It utilizes a backbone LSTM model paired with a linear layer for predicting the accuracy score.
Each input code prompt is tokenized using the BPE tokenizer, encoded via the LSTM model, and then passed through the linear layer for score prediction.
Its training is also conducted with MSE-based loss optimization.

\smallskip \noindent \textbf{Machine Learning (ML) Algorithms}: 
For an even more lightweight alternative, we explored ML algorithms, particularly Linear Regression (\textbf{LR}) and Adaboost (\textbf{ADA})~\cite{Freund1997ADG}.
LR is a classical and basic ML algorithm that learns the weights for each feature to make predictions, while ADA is an ensemble method that combines the predictions of multiple weak learners to create a strong learner.
In our implementation, the base learners of ADA are decision trees, and each tree is trained sequentially, with each new tree giving more weight to the miscalculated samples from the previous trees.
These algorithms differ from deep learning models in how they process the code prompt.
More specifically, they use feature vectors instead of tokens.
We utilized the widely used TF-IDF technique for feature extraction, converting code prompts into weighted vector representations.
Given the features obtained through TF-IDF, the ML estimators predict the accuracy score.
The training for both ML-based estimators minimizes the MSE loss, similar to the deep learning models.
}

\section{Experimental setup}
\label{sec:setup}

\su{
This section introduces the research questions, large code models, datasets, evaluation metrics, and implementation details of our experiments.
The goal of our experiments is to evaluate the accuracy and efficiency of each implemented estimator, which further demonstrates the feasibility of \mechanism.
}
Therefore, we propose three research questions and design several experiments to answer each of them:

\begin{itemize}[leftmargin=*]

    \item \textbf{RQ1:} \su{Can the implemented estimators accurately predict the quality of code completion for each prompt?}

   \item \textbf{RQ2:} How is the development productivity under the help of LCMs with \mechanism?

   \item \textbf{RQ3:} Compared with vanilla LCMs, how cost-friendly are those with \mechanism during inference stage?

\end{itemize}

\subsection{Large Code Models}
In the evaluation, we adopt two popular LCMs from our research community, \su{StarCoder and CodeGen2}, to fulfill the code completion tasks for a specific programming language.
Later, we evaluate how well \su{each estimator} can predict the quality of completions generated by them.

\begin{itemize}[leftmargin=*]
\su{\item \textbf{StarCoder (7B)}:
StarCoder~\cite{Li2023StarCoderMT} is a recently released LCM under the architecture of the GPT-2 model with multi-query attention and Fill-in-the-Middle training objective~\cite{bavarian2022efficient}.
It is pre-trained on 1 trillion tokens collected from permissively-licensed source code files.
We adopt its 7B version as the target LCM to be enhanced by \mechanism, which is the largest size that can be deployed on our server.
}

\su{\item \textbf{CodeGen2 (7B)}: CodeGen2~\cite{Nijkamp2023CodeGen2LF} is a subsequent version of CodeGen~\cite{Nijkamp2022CodeGenAO} with the same architecture but larger scale and optimized training methods.
It is trained on source code datasets with 400 billion tokens.
Similar to StarCoder, we utilize its 7B version for our experiments.
}

\end{itemize}

\subsection{Datasets}
\label{sec:exp_datasets}
We focus on the Java and Python programming languages in our experiments which have been extensively studied in code-related DL tasks.
Theoretically, the estimators are generic and could be applied to any other general programming language.
\su{\mechanism is to prevent unhelpful completions in the practice of code completion, where the input to the LCMs is partial source code of diverse formats.
Therefore, we use code snippet datasets to craft the partial code instead of adopting code generation benchmarks, such as HumanEval~\cite{Chen2021EvaluatingLL}, that are based on natural language requirements.}
The training datasets for Java and Python are listed as follows.
\begin{itemize}[leftmargin=*]
\item \textbf{Java}:
We use COFIC~\cite{Sun2022OnTI} as our Java dataset.
It is a collection of function definitions from Java repositories on Github, containing 849,984 functions, each of which is used as a code snippet.
\su{
Different from raw code datasets, COFIC is constructed with a powerful comment filter that filters out code snippets containing low-quality code comments, thus mitigating the potential bias during evaluation.}
We split the dataset into three portions: 90\% for training, 5\% for validating, and 5\% for testing.

\item \textbf{Python}:
The Python dataset is CodeSearchNet (CSN)~\cite{Husain2019CodeSearchNetCE}, \su{also a collection of functions definitions from Github.}
It has been pre-split where the train, valid, and test split respectively contain 453,772, 23,107, and 22,175 code functions.
Similarly, each function is used as a code snippet.
\end{itemize}

\begin{table}[t]
\caption{\su{The statistics of HumanAccept.}}
\centering
    \begin{tabular}{|c|c|c|c|} 
\hline
\textbf{Subsets} & \textbf{\#} & \textbf{Acceptance Rate} & \textbf{BLEU} \\ 
\hline
CodeGen2-Python & 378 & 25.30\% & 0.408 \\
CodeGen2-Java & 381 & 30.80\% & 0.48 \\
StarCoder-Python & 378 & 22.50\% & 0.394 \\
StarCoder-Java & 381 & 31.10\% & 0.483 \\
\hline
\end{tabular}
\label{tab:humanaccept}
\end{table}

\su{The training portion of the two datasets is used to train the estimators, the validating portion is to tune the hyper-parameters and monitor the model performance during training, and the test portion is used to evaluate the performance of estimator models.
Each sample in both Python and Java datasets is split into two parts at a random position: the initial part forms the code prompt for a code completion request, and the latter part provides the correct completion as ground truth.
After the splitting, the code prompts of Python and Java datasets respectively contain 271.0 and 347.5 characters.
Additionally, we have developed a new evaluation benchmark, HumanAccept, specifically designed to evaluate the estimators' proficiency in preventing unhelpful code completions.
\begin{itemize}[leftmargin=*]
\item \textbf{HumanAccept}:
This human-annotated dataset is designed to supply acceptance information for code completions produced by two specific LCMs, CodeGen2 and StarCoder, across the Python and Java programming languages.
It comprises four subsets tailored to each combination of LCM and programming language: CodeGen2-Python, CodeGen2-Java, StarCoder-Python, and StarCoder-Java.
Each dataset entry contains a code prompt, the generated code completion from the respective LCM, and an annotation indicating whether the completion is accepted.
The code prompt is randomly sampled from the test set of the corresponding programming language, using~\Cref{eq:1} to compute the sampling size.
The code prompt sampling and subsequent human annotation follow the same process as the annotation for identifying unhelpful code completions introduced in~\Cref{sec:prompts}, i.e., the two authors separately judge whether each completion should be accepted based on the same criteria and discuss to reconcile all disagreements.
The Cohen’s Kappa agreement levels between the two authors are respectively 0.75 for CodeGen2-Python, 0.82 for CodeGen2-Java, 0.71 for StarCoder-Python, and 0.84 for StarCoder-Java, all demonstrating the substantial agreement between the two authors.
The statistics of HumanAccept are reported in~\Cref{tab:humanaccept}.
\end{itemize}

}

\subsection{Evaluation Metrics}
Four popular metrics are adopted in our experiments:
\begin{itemize}[leftmargin=*]
\item \textbf{BLEU}:
BLEU~\cite{Papineni2002BleuAM} is widely adopted to measure the accuracy of LCMs
by counting the matched n-grams between an LCM-generated code and its ground truth completion for each prompt.
Furthermore, following Lin et al.~\cite{lin-och-2004-orange}, we smooth the BLEU to give non-zero credits to short completions.

\item \textbf{CrystalBLEU (CBLEU)}:
CrystalBLEU~\cite{Eghbali2022CrystalBLEUPA} is a variant of the BLEU metric, specifically designed for measuring the similarity between source code.
It ignores the trivially shared n-grams in the code corpus, hence reducing the noises of BLEU.


\item \su{\textbf{Acceptance Rate (Accept)}:
Acceptance Rate indicates the proportion of completions that are marked as accepted, i.e., the number of accepted completions divided by the number of all completions.
As investigated by ~\cite{Ziegler2022ProductivityAO}, the acceptance rate is the best metric for measuring the perceived productivity of developers, compared with tens of alternative measures in the practice of code completion systems.}

\item \su{\textbf{Precision:} 
Precision is the proportion of predicted samples that are correctly classified.
We use it to measure the accuracy of the estimator when a fixed proportion of code prompts is identified as low-return.
As a lightweight estimator, it is enough to produce a positive impact in terms of efficiency or productivity, even if only a small proportion of low-return prompts can be precisely rejected.}

\item \textbf{Floating point operations (FLOPs)}:
FLOPs is the number of floating-point operations needed for executing an instance.
\su{1 MetaFLOPs,  GitaFLOPs, and TeraFLOPS, respectively, indicate one million ($10^6$), billion ($10^9$), and trillion ($10^{12}$) FLOPs.}
It is a popular device-agnostic metric for measuring the efficiency of DL models~\cite{Strubell2019EnergyAP, Desislavov2021ComputeAE} from which
the concrete energy consumption can be deduced based on the computational efficiency of GPU devices.

\end{itemize}

\subsection{Implementation Details}
\su{Both LCMs, StarCoder and CodeGen2, are implemented using Huggingface Transformers and loaded in 4-bit quantization for inference to accommodate our hardware constraints.
The maximum number of input tokens and generated tokens for one code completion task is set to 256 and 10, which is enough for LCMs to understand the code prompt and provide a full-line code completion suggestion.
The experiments are performed on a server running an Ubuntu 18.04 operation system with an AMD EPYC 7643 CPU {(48 Core)}, 512GB RAM, and four RTX A5000 GPUs {(24GB RAM)}.
It is noteworthy that the different combinations of LCMs, programming languages, and evaluation metrics yield different training data for the estimator, and a separate estimator is trained for each combination following the same design introduced in~\Cref{sec:estimator_design}.}

\section{Experimental Results}
\label{sec:results}

In this section, we report our experimental results and answer the three research questions.

\subsection{RQ1: Accuracy of estimators}
\label{sec:rq1}




\begin{table}[t]
\caption{The Precision of each estimator on the human-annotated evaluation set.}
\small
\setlength{\tabcolsep}{2pt}
\centering
    \begin{tabular}{|c|c|c|c|c|c|c|c|} 
\hline
\textbf{Model} & \textbf{Language} & \textbf{Baseline} & \textbf{Estimator} & \textbf{Bottom 5\%} & \textbf{Bottom 10\%} & \textbf{Bottom 20\%} & \textbf{Bottom 50\%} \\ 
\hline
\multirow{10}{*}{CodeGen2} & \multirow{5}{*}{Python} & \multirow{5}{*}{74.7\%} & LR & 68.4\% & 68.4\% & 73.7\% & 77.0\% \\
 &  &  & ADA & 84.2\% & 89.5\% & 88.2\% & 79.1\% \\
 &  &  & LSTM & 94.7\% & 84.2\% & 81.6\% & 75.9\% \\
 &  &  & ET & \textbf{100.0\%} & \textbf{97.4\%} & \textbf{93.4\%} & \textbf{83.8\%} \\
 &  &  & DT & \textbf{100.0\%} & \textbf{97.4\%} & \textbf{93.4\%} & 81.2\% \\ 
\cline{2-8}
 & \multirow{5}{*}{Java} & \multirow{5}{*}{69.2\%} & LR & 73.7\% & 84.2\% & 82.9\% & 81.2\% \\
 &  &  & ADA & \textbf{100.0\%} & 97.4\% & 90.8\% & 83.8\% \\
 &  &  & LSTM & 84.2\% & 78.9\% & 78.9\% & 79.6\% \\
 &  &  & ET & \textbf{100.0\%} & \textbf{100.0\%} & \textbf{97.4\%} & \textbf{85.9\%} \\
 &  &  & DT & 94.7\% & 92.1\% & 90.8\% & 80.1\% \\ 
\hline
\multirow{10}{*}{StarCoder} & \multirow{5}{*}{Python} & \multirow{5}{*}{77.5\%} & LR & 73.7\% & 65.8\% & 77.6\% & 80.1\% \\
 &  &  & ADA & \textbf{100.0\%} & \textbf{97.4\%} & 94.7\% & 85.9\% \\
 &  &  & LSTM & 89.5\% & 86.8\% & 84.2\% & 81.7\% \\
 &  &  & ET & \textbf{100.0\%} & \textbf{97.4\%} & \textbf{97.4\%} & \textbf{87.4\%} \\
 &  &  & DT & \textbf{100.0\%} & \textbf{97.4\%} & 94.7\% & 86.9\% \\ 
\cline{2-8}
 & \multirow{5}{*}{Java} & \multirow{5}{*}{68.9\%} & LR & 68.4\% & 73.7\% & 80.3\% & 79.1\% \\
 &  &  & ADA & \textbf{100.0\%} & 97.4\% & \textbf{94.7\%} & 81.2\% \\
 &  &  & LSTM & 94.7\% & 86.8\% & 80.3\% & 71.2\% \\
 &  &  & ET & \textbf{100.0\%} & \textbf{100.0\%} & 92.1\% & \textbf{87.4\%} \\
 &  &  & DT & 89.5\% & 92.1\% & 92.1\% & 85.9\% \\
\hline
\end{tabular}
\label{tab:rq1}
\end{table}

\su{To answer RQ1, we evaluate how accurately various estimators predict the completion quality of different LCMs within the \mechanism framework.
As mentioned in~\Cref{sec:DCE}, training an estimator model requires a code dataset, a target LCM, and an accuracy metric.
For our experiments, we prepared two code datasets (i.e., Python and Java), two LCMs (i.e., CodeGen2 and StarCoder), and two metrics (i.e., BLEU and CBLEU) to evaluate the five estimators (ET, DT, LSTM, LR, and ADA).
Correspondingly, for each estimator, we trained eight models under these settings and applied them to estimate the completion quality of the corresponding LCM.
We use each estimator to produce an estimation score for each prompt in the HumanAccept.
In practice, \mechanism is to reject the low-rated prompts; therefore, we care more about the accuracy of the estimator when giving low scores.
We thus respectively compute the precision when rejecting the lowest 5\%, 10\%, 20\%, and 50\% of prompts by their estimation scores, namely, their scores are less than a threshold decided by the distribution.
We also compute a baseline precision for each setting, which indicates the expected precision of randomly rejecting prompts at a 50\% chance.

The results of the estimators under different settings are reported in \Cref{tab:rq1}.
For simplicity, we report the higher precision between BLEU-trained and CBLEU-trained estimators in this paper, and the full results are available in our artifacts.
Firstly, except for the LR estimator, all other estimators demonstrate positive results across most settings, with a notable improvement over the baseline precision.
For example, when discarding the lowest 20\% of prompts, on average of all settings, ADA, LSTM, ET, and DT estimators achieve precisions of 92.1\%, 81.3\%, 95.1\%, and 92.8\%, respectively, compared to a 72.6\% baseline precision.
Remarkably, ET achieves an exceptional 97.4\% precision with CodeGen2 in Java, illustrating the \mechanism's effectiveness in utilizing learning-based estimations for code completion quality.
Among the evaluated estimators, ADA, ET, and DT show commendable performance, with ET, an encoder-only transformer model, consistently outperforming others.
For example, ET rejects 10\% of prompts with a 98.7\% precision on average, 3.3\% higher than the next best estimator.
ADA also demonstrates high precisions, particularly in scenarios with fewer rejections (5\% or 10\%).
It can reject 5\% of prompts for StarCoder in both Java and Python with a 100\% precision, highlighting the potential of machine learning algorithms as efficient estimators.
Our analysis also indicates BLEU as a more suitable accuracy metric than CBLEU for code completion quality assessments. 
CBLEU cannot correctly handle the code prompt that needs syntax symbols for code completion.
Take ET for example.
BLEU-trained estimators outperform CBLEU-trained counterparts in 14 out of 16 pairs due to CBLEU's omission of common syntax symbols for mitigating the noises of exact-match-based metrics.
Finally, we also observe that the performance of the estimators varies significantly with the setting.
For example, ADA achieves an 84.2\% Precision in rejecting 5\% of prompts for CodeGen2 on Python, while it can reach 100.0\% Precision for StarCoder under the same other settings.
This variability suggests the importance of customizing estimator configurations based on LCM's performance, advocating for preliminary testing on relevant datasets.
}

\begin{tcolorbox}[size=title]
{\textbf{Answer to RQ1:}}
\su{Learning-based estimators, including both machine learning and deep learning models, can reject a considerable amount of low-return prompts with extremely high precision, such as the 97.4\% Precision achieved by ET when rejecting 20\% of the code prompts for Starcoder (Python), which demonstrates the feasibility of \mechanism.}
\end{tcolorbox}

\subsection{RQ2: Development productivity under the help of LCMs with \mechanism}
\label{sec:rq2}

\begin{table}[t]
\caption{The BLEU and acceptance rate of the completions generated by the LCM with each estimator. Ret. and Rej. are respectively short for Retained and Rejected.}
\small
\setlength{\tabcolsep}{2.5pt}
\centering
    \begin{tabular}{|c|c|c|c|c|cc|cc|cc|cc|} 
\hline
\multirow{2}{*}{\textbf{Meric}} & \multirow{2}{*}{\textbf{Model}} & \multirow{2}{*}{\textbf{Language}} & \multirow{2}{*}{\textbf{Baseline}} & \multirow{2}{*}{\textbf{Est.}} & \multicolumn{2}{c|}{\textbf{Reject 5\%}} & \multicolumn{2}{c|}{\textbf{Reject 10\%}} & \multicolumn{2}{c|}{\textbf{Reject 20\%}} & \multicolumn{2}{c|}{\textbf{Reject 50\%}} \\ 
\cline{6-13}
 &  &  &  &  & \textbf{Rej.} & \textbf{Ret.} & \textbf{Rej.} & \textbf{Ret.} & \textbf{Rej.} & \textbf{Ret.} & \textbf{Rej.} & \textbf{Ret.} \\ 
\hline
\multirow{8}{*}{BLEU} & \multirow{4}{*}{CodeGen2} & \multirow{2}{*}{Python} & \multirow{2}{*}{0.408} & ADA & 0.232 & 0.418 & 0.245 & 0.426 & 0.275 & 0.442 & 0.341 & 0.475 \\
 &  &  &  & ET & 0.209 & 0.419 & 0.222 & 0.429 & 0.251 & 0.448 & 0.324 & 0.493 \\ 
\cline{3-13}
 &  & \multirow{2}{*}{Java} & \multirow{2}{*}{0.480} & ADA & 0.204 & 0.494 & 0.238 & 0.507 & 0.256 & 0.536 & 0.370 & 0.589 \\
 &  &  &  & ET & 0.104 & 0.500 & 0.149 & 0.517 & 0.197 & 0.551 & 0.327 & 0.633 \\ 
\cline{2-13}
 & \multirow{4}{*}{StarCoder} & \multirow{2}{*}{Python} & \multirow{2}{*}{0.394} & ADA & 0.191 & 0.405 & 0.208 & 0.415 & 0.237 & 0.433 & 0.312 & 0.476 \\
 &  &  &  & ET & 0.176 & 0.405 & 0.196 & 0.416 & 0.219 & 0.438 & 0.298 & 0.490 \\ 
\cline{3-13}
 &  & \multirow{2}{*}{Java} & \multirow{2}{*}{0.483} & ADA & 0.180 & 0.499 & 0.219 & 0.512 & 0.252 & 0.540 & 0.371 & 0.594 \\
 &  &  &  & ET & 0.090 & 0.503 & 0.136 & 0.521 & 0.184 & 0.557 & 0.328 & 0.638 \\ 
\hline
\multirow{8}{*}{Accept} & \multirow{4}{*}{CodeGen2} & \multirow{2}{*}{Python} & \multirow{2}{*}{25.3\%} & ADA & 15.8\% & 25.8\% & 10.5\% & 27.0\% & 11.8\% & 28.7\% & 20.9\% & 29.7\% \\
 &  &  &  & ET & 0.0\% & 26.6\% & 2.6\% & 27.8\% & 6.6\% & 30.0\% & 16.2\% & 34.4\% \\ 
\cline{3-13}
 &  & \multirow{2}{*}{Java} & \multirow{2}{*}{30.8\%} & ADA & 0.0\% & 32.4\% & 2.6\% & 33.9\% & 9.2\% & 36.2\% & 16.2\% & 45.3\% \\
 &  &  &  & ET & 0.0\% & 32.4\% & 0.0\% & 34.2\% & 2.6\% & 37.8\% & 14.1\% & 47.4\% \\ 
\cline{2-13}
 & \multirow{4}{*}{StarCoder} & \multirow{2}{*}{Python} & \multirow{2}{*}{22.5\%} & ADA & 0.0\% & 23.6\% & 2.6\% & 24.6\% & 5.3\% & 26.7\% & 14.1\% & 30.7\% \\
 &  &  &  & ET & 0.0\% & 23.6\% & 2.6\% & 24.6\% & 2.6\% & 27.4\% & 12.6\% & 32.3\% \\ 
\cline{3-13}
 &  & \multirow{2}{*}{Java} & \multirow{2}{*}{31.1\%} & ADA & 0.0\% & 32.7\% & 2.6\% & 34.2\% & 5.3\% & 37.5\% & 18.8\% & 43.2\% \\
 &  &  &  & ET & 0.0\% & 32.7\% & 0.0\% & 34.5\% & 7.9\% & 36.8\% & 12.6\% & 49.5\% \\
\hline
\end{tabular}
\label{tab:rq2}
\end{table}
\su{
To answer RQ2, we focus on assessing the impact of the \mechanism mechanism on developers' productivity.
Specifically, we evaluate the effectiveness of estimators in preventing unhelpful code completions through two primary metrics: the BLEU score across the entire test set and the Acceptance Rate within the HumanAccept dataset. 
BLEU is an exact-match-based metric indicating the correctness of the completions, while the Acceptance Rate is measured from the view of developers to profile the helpfulness of the completions.
Both types of metrics are essential for comprehensively understanding the development productivity boosted by the completions under \mechanism.
These evaluations involve two LCMs, CodeGen2 and StarCoder, across two programming languages, Python and Java.
For each code prompt of the test set and HumanAccept, we apply the LCM to generate a piece of code completion, which is then selectively rejected by the estimators based on pre-defined thresholds.
Each estimator is trained with BLEU metric and the threshold is set to reject 5\%, 10\%, 20\%, and 50\% prompts.
Additionally, we compute a baseline for both BLEU scores and Acceptance Rates by evaluating the LCM outputs without any estimator intervention, offering a direct comparison to understand the added value of the estimator application.

For the sake of brevity, this paper discusses the performance of the two leading estimators identified in RQ1: ADA and ET, which represent the standout machine learning and deep learning approaches, 
The broader results, including those of other estimators, are detailed within our supplemental artifacts.
We report the average performance of the retained and rejected groups in ~\Cref{tab:rq2}
According to the results, employing estimators significantly improves the BLEU and Acceptance Rate of retained completions over baseline metrics.
For instance, ET's rejection of 5\%, 10\%, 20\%, and 50\% of prompts elevates the average BLEU from 0.441 to respective heights of 0.457, 0.471, 0.498, and 0.563, while similarly raising the Acceptance Rate from 27.4\% to 28.8\%, 30.3\%, 33.0\%, and 40.9\%.
It manifests the effectiveness of \mechanism in blocking low-return prompts in practice, as evidenced by the low average Acceptance Rates of rejected prompts, 1.3\% by ET and 4.6\% by ADA at a 10\% rejection rate. 
Moreover, ET shows a more effective and stable performance than ADA in enhancing the quality of retained code completions.
For example, with 50\% of prompts rejected, ET boosts the average Acceptance Rate to 40.9\%, compared to ADA's 37.2\%.
This analysis also highlights a critical trade-off regarding the setting of the thresholds.
All Java prompts from CodeGen2 discarded by ET at a 10\% rejection rate are deemed unhelpful (0\% Acceptance Rate), yet this shifts to 14.1\% deemed helpful at a 50\% rejection rate. 
All the Java prompts from CodeGen2 rejected by ET at 10\% are unhelpful (i.e., 0\% acceptance rate), but 14.1\% of rejected prompts are considered helpful when the rejection rate increases to 50\%.
This variance underscores the flexibility in setting estimator thresholds to align with either conservative or aggressive preferences, tailoring to specific system requirements or user expectations.
}

\begin{tcolorbox}[size=title]
{\textbf{Answer to RQ2:}
\su{On average of all settings when rejecting bottom 20\% prompts, only 4.9\% of those rejected by the ET are considered acceptable.
This highlights the efficacy of the estimators in filtering out low-return prompts in practice. 
Consequently, the acceptance rate of the displayed code completions rises from 27.4\% to 33.0\%, showcasing a significant improvement in productivity.}}
\end{tcolorbox}

\subsection{RQ3: Cost-friendliness of \estimator}
\label{sec:rq3}

\su{To answer RQ3, we assess the cost-friendliness of various estimators, LR, ADA, LSTM, DT, and ET, relative to two LCMs, CodeGen2 and StarCoder.
The cost-friendliness is evaluated using the computational expense required by both LCMs and estimators to process a prompt comprising 256 tokens, where LCMs generate code completions (of 10, 20, and 50 new tokens) while estimators provide a corresponding estimation score. 
We compute the total number of floating point operations (represented by FLOPs) and running time during the code completion or score prediction as the indicator of computational cost.
Notably, in practice, estimators could be deployed on user devices, potentially offloading the computational demand from LCM servers, which offers a practical resource-saving benefit.
}

\begin{table}[t]
\caption{\su{Scale and inference cost of estimators and LCMs. $m$ is the number of tokens in a piece of code completion generated by the LCM. The parameters are non-embedding parameters.}}
\small
\setlength{\tabcolsep}{2.5pt}
\centering
    \begin{tabular}{|c|c|c|c|c|c|c|c|c|} 
\hline
\multirow{2}{*}{\textbf{Type}} & \multirow{2}{*}{\textbf{Models}} & \multirow{2}{*}{\textbf{Parameters}} & \multicolumn{3}{c|}{\textbf{FLOPs}} & \multicolumn{3}{c|}{\textbf{Time}} \\ 
\cline{4-9}
 &  &  & \textbf{m = 10} & \textbf{m = 20} & \textbf{m = 50} & \textbf{m = 10} & \textbf{m = 20~} & \textbf{m = 50} \\ 
\hline
\multirow{2}{*}{LCM} & CodeGen2 & 6.7 B & 3.42 T & 3.42 T & 3.44 T & 3.9 s & 7.1 s & 17.1 s \\
 & StarCoder & 6.7 B & 3.75 T & 3.76 T & 3.78 T & 1.7 s & 7.4 s & 19.0 s \\ 
\hline
\multirow{5}{*}{Estimator} & LR & - & \multicolumn{3}{c|}{ \textless 1 M} & \multicolumn{3}{c|}{0.1 ms} \\
 & ADA & - & \multicolumn{3}{c|}{ \textless 1 M} & \multicolumn{3}{c|}{0.1 ms} \\
 & LSTM & 0.3 M & \multicolumn{3}{c|}{0.27 G} & \multicolumn{3}{c|}{3.9 ms} \\
 & DT & 7.9 M & \multicolumn{3}{c|}{3.80 G} & \multicolumn{3}{c|}{5.2 ms} \\
 & ET & 7.0 M & \multicolumn{3}{c|}{3.83 G} & \multicolumn{3}{c|}{5.1 ms} \\
\hline
\end{tabular}
\label{tab:rq3}
\end{table}

\su{
We report the computational scale and cost for each estimator and LCM in \Cref{tab:rq3}.
Compared to the LCMs, which operate on a scale of billions of parameters, all evaluated estimators require minimal computational resources and time. 
Specifically, every estimator is capable of performing an estimation within 6 milliseconds, consuming less than 4 GFLOPs.
In contrast, CodeGen2 and StarCoder demand considerably more computational power, needing between 3.42 to 3.78 TFLOPs and taking 1.7 to 19.0 seconds to generate code completions of varying lengths
This discrepancy highlights the estimators' potential to conserve computational resources, especially since rejecting even a tiny fraction of prompts (less than 1\%) results in a net positive saving of computing resources for both LCMs.

Among the estimators, the machine learning-based LR and ADA stand out for their speed, completing tasks in less than 0.1 milliseconds, and do not require a GPU for deployment, making them ideal for running on user devices. 
While deep-learning-based estimators necessitate a GPU environment, they still operate within acceptable efficiency bounds and are capable of delivering more precise estimations.
This balance between efficiency and accuracy underscores the cost-friendly advantage of incorporating estimators into the \mechanism system, providing a scalable solution that enhances the overall computational economy.
}

\begin{tcolorbox}[size=title]
{\textbf{Answer to RQ3:}}
\su{Compared with large code models, the estimators run at a very low cost, at most 3.83 GFLOPs and 5.2 ms running time. 
For LCMs,  more resources are saved when declining more requests; \mechanism can help save GFLOPs and running time when rejecting no more than 1\% of the requests.}
\end{tcolorbox}

\section{Related Work}
\label{sec:related}
\subsection{Neural Code Completion}
The deep learning techniques used for neural code completion are constantly evolving.
At the very beginning, the models used in code completion studies were dominated by train-from-scratch models, such as Recurrent Neural Networks (RNN) and their variants~\cite{Li2018CodeCW, Svyatkovskiy2019PythiaAC}.
For example, Li et al.~\cite{Li2018CodeCW} propose to enhance the LSTM model with a pointer component to solve the out-of-vocabulary problem in code completion systems.
Such models are lightweight and can be easily deployed on users' devices.
With the rise of a new paradigm in deep learning, i.e., pre-training and fine-tuning, the focus of code completion studies has shifted to transformer-based pre-trained models, including the encoder-decoder~\cite{Li2022CompetitionLevelCG} and decoder-only~\cite{Chen2021EvaluatingLL,Nijkamp2022ACP} transformer architecture.
The pre-trained models with a large number of trainable parameters, usually over hundreds of millions, are known as Large Language Models (LLM), where LCM is the variant of LLM in the domain of source code.
LCM refers to the large model that is specifically trained with a large-scale code corpus, e.g., by pre-training a base LLM~\cite{Feng2020CodeBERTAP, Wang2021CodeT5IU}, or fine-tuning a pre-trained LLM~\cite{Chen2021EvaluatingLL}, to generate code.
Recent studies~\cite{Lu2021CodeXGLUEAM, Chen2021EvaluatingLL} have empirically demonstrated the state-of-the-art performance of LCMs across multiple code-related tasks.
\su{For example, a recently released LLM, AlphaCode2~\cite{alphacode2}, is reported to perform better than 85\% of human participants in a programming competition.
Many LCMs are thus proposed in the research field of code generation~\cite{Li2023StarCoderMT, Nijkamp2022CodeGenAO, Luo2023WizardCoderEC, Hou2023LargeLM, Zheng2023ASO}.
Such models also appear in commercial code completion systems, such as Github Copilot~\cite{copilot} and Amazon CodeWhisperer~\cite{codewhisperer}.}
Generally speaking, the more parameters in a model, the greater the computational resources required for inference.
Small models are lightweight enough to run on local devices, while large models have to be deployed on a large number of high-performance GPU servers, providing services to clients via remote APIs.
As a result, using LCMs is costly, both financially, due to the cost of servers, and environmentally, due to the carbon footprint required to fuel the hardware devices~\cite{Strubell2019EnergyAP,Desislavov2021ComputeAE}.
This greatly limits the commercialization and further research of LCMs.
Researchers have proposed many methods to reduce such costs, such as weight sharing~\cite{Rothe2020LeveragingPC}, pruning~\cite{See2016CompressionON} and knowledge distillation~\cite{Gou2021KnowledgeDA}.
However, it is still a challenging problem that calls for community efforts to build an environment-friendly and sustainable future for LCMs.

\subsection{Performance Estimation for Neural Models}
\su{Performance estimation is to estimate the performance of a neural model without actually training or testing the model, which is a classic task in machine learning~\cite{ljung2002prediction, molinaro2005prediction, kohavi1995automatic}.
It is widely applied to various domains when it comes to neural models.
For example, Xia et al.~\cite{Xia2020PredictingPF} predict the performance of translation models given their experimental settings to bypass the computation restriction.
Wei et al.~\cite{Wei2021MetalearningHP} estimate the performance of the hyperparameters of neural models to expedite the automated hyperparameter optimization for meta-learning.
Alshubaily~\cite{Alshubaily2021EfficientNA} accelerates the fitness evaluation during the neural architecture search with an end-to-end performance predictor.
FilterForward~\cite{Canel2019ScalingVA} trains a binary classifier to determine whether to transmit the input image to the server.
Inspired by the successes in these domains, we adopt the idea of performance estimation to mitigate the low-return prompts for LCM-based code completion systems, which is a brand-new and important field with a number of new challenges to address.
}

\section{Discussion}
\label{sec:threats}
\subsection{Threats to validity}
\smallskip
\noindent\textbf{Limited Experiments}.
Limited by our computational resources, we only investigate the effectiveness of learning-based estimators on two LCMs.
In theory, our approach is capable of any LCMs.
Yet, the effectiveness of the estimators in more LCMs has not been experimentally studied.
Besides, though being sampled with a statistically decided sample size, the small-scale groups of the benchmark, HumanAccept, may introduce bias to the evaluation of our approach.

\smallskip
\noindent\textbf{Threshold}.
\mechanism rejects the code prompts according to their estimation scores with a pre-set thresholds.
Though we have explored several threshold settings in~\Cref{sec:rq1}, it is still a non-trivial task to set a proper threshold for an LCM-based code completion system considering its complex production environment.
Further studies on the settings of the threshold for the optimal effects on preventing unhelpful prompts are desired.

\smallskip
\su{\noindent\textbf{Manual Analysis}
The manual analysis and annotations were conducted by two of our authors. Despite their expertise, human annotators may introduce inherent biases in their judgment, potentially affecting the evaluation results and conclusions.
To mitigate this threat, the two authors engaged in discussions to resolve any disagreements and reach a consensus, thereby enhancing the reliability of their judgments.
Furthermore, the involvement of only two subjects in these analyses may also introduce biases in the results.
We have made the annotated dataset publicly available for others to replicate and improve the quality of the annotations.
}

\su{
\subsection{Potential Applications}
This paper concentrates on enhancing the efficacy and helpfulness of LCM-based code completion systems.
Beyond these objectives, \mechanism exhibits significant potential in addressing a range of issues.
For example, \mechanism could serve as a defensive measure against adversarial attacks on LCMs, such as ~\cite{Yang2022NaturalAF, yang2024stealthy}, since adversarial examples may exhibit patterns akin to those of low-return prompts, which lead to a significant decline in the quality of code completions.
A thorough empirical study is desired to explore the secondary impacts of \mechanism, presenting an interesting avenue for future research.
}

\su{
\subsection{Limitations}
Though \mechanism is demonstrated to be effective in preventing unhelpful code completion, some limitations exist in the estimators.
For example, we use automatic evaluation metrics, BLEU and CrystalBLEU, to construct a dataset for training the estimators, as manually labeling the helpfulness of completions generated for a large scale of code prompts is infeasible.
However, such exact-match-based metrics do not consider the code semantics, so may not be very reliable in measuring the quality of the code completions.
The shortcomings of such metrics have been investigated by the research community~\cite{Evtikhiev2022OutOT,Roy2021ReassessingAE}, but still remain a significant challenge.
For the estimators, the acceptance of a piece of code completion is the most desirable indicator for the quality.
Though we have annotated a small group of code completions for the evaluation, it is far from enough to serve as the training dataset.
A large-scale training dataset needs to be collected from a production environment, which is beyond our capabilities.
We, therefore, call for more industrial participation to facilitate further research on this topic.
}

\section{Conclusion}
\label{sec:conclusion}
We investigated the unhelpful code completion problem in the context of neural code completion due to low-return prompts,
which not only poses a threat to the developer productivity but also leads to a huge waste of computational resources.
We conducted an empirical study via manual inspection to analyze the patterns of low-return prompts and discovered four patterns that are frequently observed to cause unhelpful code completions.
Based on this finding, we proposed the first early-rejection mechanism, \mechanism, to turn down lower-return prompts by estimating the quality of their completions without activating the LCM.
Furthermore, we design and implement a range of lightweight learning-based estimators.
The experimental results show that the estimators can accurately block low-return prompts based on various threshold settings, which demonstrates the feasibility of the mechanism.
In the future, we seek to design estimators that are more effective and lightweight to facilitate the industrial applications of our mechanism.

\begin{acks}
This research / project is supported by the 
National Natural Science Foundation of China (62072309),
CAS Project for Young Scientists in Basic Research (YSBR-040), 
ISCAS New Cultivation Project (ISCAS-PYFX-202201), 
ISCAS Fundamental Research Project (ISCAS-JCZD-202302), 
the Research Foundation from NUDT (Grant No. ZK24-05),
Xiaoning Du’s Google Research Scholar Program Award,
and National Research Foundation, under its Investigatorship Grant (NRF-NRFI08-2022-0002).
Any opinions, findings and conclusions or recommendations expressed in this material are those of the author(s) and do not reflect the views of National Research Foundation, Singapore.
\end{acks}

\balance

\bibliographystyle{ACM-Reference-Format}
\bibliography{reference}
\end{document}